# Solution to the King's Problem in prime power dimensions

P.K.Aravind
Physics Department
Worcester Polytechnic Institute
Worcester, MA 01609
(Email: paravind@wpi.edu)

ABSTRACT

It is shown how to ascertain the values of a complete set of mutually complementary observables of a prime power degree of freedom by generalizing the solution in prime dimensions given by Englert and Aharonov [Phys. Lett. A284, 1-5 (2001)].

# 1. Introduction

The *King's Problem*[1] is the following: A physicist is trapped on an island ruled by a mean king who promises to set her free if she can give him the answer to the following puzzle. The physicist is asked to prepare a $d-$state quantum system in any state of her choosing and give it to the king, who measures one of several sets of mutually unbiased observables (this term will be defined below) on it. Following this, the physicist is allowed to make a control measurement on the system, as well as any other systems it may have been coupled to in the preparation phase. The king then reveals which set of observables he measured and the physicist is required to predict correctly all the eigenvalues he found. A special case of this problem was first introduced and solved in Ref.[2] for the case of $d=2$: there the king is given a spin-1/2 particle (or qubit) and allowed to measure one of the spin components $\sigma_x, \sigma_y$ or $\sigma_z$ on it. Some variants of this basic problem were discussed in Refs.[3]-[5]. Then a solution to the problem for $d=3$ was presented in Ref.[6], following which a solution for arbitrary prime $d$ was presented in Ref.[1]. The purpose of this paper is to generalize the solution in Ref.[1] to arbitrary prime power dimensions.

Let $A$ and $B$ be two observables of a $d-$state quantum system with orthonormal eigenstates $\{|\alpha_i\rangle\}$ and $\{|\beta_i\rangle\}$, respectively. These observables (and their eigenstates) are said to be mutually unbiased[7] if the inner products of all pairs of eigenstates across the two bases have the same magnitude, i.e. if $|\langle\beta_i|\alpha_j\rangle|=1/\sqrt{d}$ for all $i$ and $j$. The observables are also sometimes spoken of as being "mutually complementary" or "maximally non-commutative"[8] because, given any eigenstate of one, one is completely uncertain about the eigenvalue one will get upon a measurement of the other. The notion of mutual unbiasedness extends naturally from two to a larger number of bases: any set of bases is said to be mutually unbiased if all pairs of bases within them are mutually unbiased. It can be shown[7] that a $d-$dimensional Hilbert space has a maximum of $d+1$ mutually unbiased bases. An explicit construction of a maximal set of mutually unbiased bases in prime dimensions was given by Ivanovic[9], following which Wootters and Fields[7] showed how to extend the construction to prime power dimensions. Several recent papers[10-12] have revisited and clarified this construction from new points of view.

A simple example of a set of mutually unbiased observables (and bases) is afforded by a qubit: the spin components $\sigma_x, \sigma_y$ and $\sigma_z$ along three orthogonal directions constitute a set of mutually unbiased observables and their orthonormal eigenstates a set of mutually unbiased bases. An important additional point can be made: the observables $\sigma_x, \sigma_y$ and $\sigma_z$ also constitute a *complete* and *minimal* set of observables for a qubit. They constitute a *complete* set because a knowledge of their expectation values in an arbitrary state of the qubit, pure or mixed, serves to determine that state uniquely. And they constitute a *minimal* set because leaving even one of them out would make the state identification impossible.

The above state of affairs for a qubit leads one to ask whether a similar situation obtains for a $d-$state system. The density matrix of a $d-$state system, being a $d \times d$ hermitian matrix with unit trace, requires $d^2-1$ independent parameters to specify it completely. Alternatively, it requires $d+1$ non-degenerate observables to specify it because each observable determines $d-1$ parameters (through the $d$ probabilities for the possible outcomes, with unit sum). One can ask whether it is possible to choose these $d+1$ observables so that they are all mutually unbiased with respect to each other. That this can be done was demonstrated in Ref.[1] for prime $d$ and in Ref.[10] for prime power $d$. Following Ref.[10], one can construct $d+1$ sets of mutually unbiased observables, each consisting of $d-1$ mutually commuting observables, such that the simultaneous eigenstates of each set give rise to a system of $d+1$ mutually unbiased bases, the maximum number possible in this space.

We are now in a position to state more clearly the problem solved in this paper. The king is given a $d$-state system (with $d$ a prime power) and allowed to measure any one of $d+1$ sets of mutually unbiased observables on it. Once he has picked a set to measure, the king measures the $d-1$ commuting observables in that set and notes the eigenvalues he gets. He then tells the physicist which set he picked, and she is required to tell him all the eigenvalues he found. It turns out that not all the $d-1$ observables in each set are independent, but that they can all be generated as products of powers of a smaller subset within them. It is only this subset of observables that the king need measure, and not the entire set itself. The subset is arbitrary to some extent, but always provides a unique set of eigenvalue labels for the basis arising from that set.

In addition to its importance for the King's Problem, to be made clearer in the later sections, the construction of mutually unbiased observables/bases has other important applications: it provides the most efficient method of determining an unknown quantum state from a finite number of measurements[7], it is a crucial ingredient in certain protocols for quantum key distribution[13], and it is closely related to the notion of mutually complementary propositions that arises in the information-theoretic approach to quantum mechanics.[14]

The solution to the King's Problem given by Englert and Aharonov[1] fails for composite dimensions at two points: firstly, it fails to yield a set of mutually unbiased observables/bases and, secondly, it fails (though only narrowly) to yield the observable the physicist needs to measure in order to learn the king's results. We get around the first obstacle by using the alternative construction for mutually unbiased bases given in Ref.[10], and we overcome the second obstacle by replacing eqn.(33) of Ref.[1] by a slightly modified equation involving elements of the prime power Galois field $GF(p^n)$. Rather than describe our solution in the most general setting, we will construct it explicitly in dimensions $2^2 = 4, 2^3 = 8$ and $3^2 = 9$. Even these simple cases present several features of interest, and they also illustrate how to obtain the solution in the more general cases.

## 2. Solution in dimension d = $2^2$ = 4

The four-state system given to the king is conveniently taken to be a pair of qubits. What are the sets of mutually unbiased observables from which he is allowed to pick one to measure? One choice consists of the five sets of three observables shown in the first column of Table 1, where $X, Y, Z$ and 1 stand for the Pauli and identity operators of a qubit and the product notation $XY$ etc. is used to denote tensor products of observables for the two qubits. The simultaneous eigenstates of each set of commuting observables yield the five mutually unbiased bases shown in the table. Note that the king need measure only two out of the three observables in any set, since the third (and hence its eigenvalue) is always the product, upto a sign, of the other two. The following conventions have been used in Table 1 (and will recur later in Tables 3 and 5): (1) the eigenstates of the observables in the first row yield the standard basis; (2) the eigenstates in any row are ordered according to their eigenvalue signatures with respect to a generating set of observables (here the first two) in that row; and (3) the phases of the states in all rows after the first have been chosen so that their overlap with the standard basis state $|00\rangle$ is real positive.

What is the state the physicist should give to the king so that he can carry out his measurements? The physicist needs to use a system of four qubits, of which two are "object" qubits that she gives to the king and the other two are "ancilla" qubits that she retains in her possession. The state the physicist prepares is most conveniently expressed in terms of maximal sets of mutually unbiased bases in the object and ancilla spaces (here "maximal" refers to the requirement that one have $d+1=5$ bases, and no less). The bases in the object space are just the ones shown in Table 1, with $|m_k\rangle$ ($m = 0,1,...4$, $k = 1,...,4$) denoting the $k$-th state of the $m$-th basis. The bases in the ancilla space can be chosen to be identical to the object bases, but it turns out to be more convenient to choose them so that $\langle 0_j | m_k \rangle = \langle \bar{m}_k | \bar{0}_j \rangle$, where the overbars denote ancilla states. The foregoing condition can be met by taking $|\bar{0}_j\rangle = |0_j\rangle$ and choosing $|\bar{m}_k\rangle$ to be the complex conjugate of $|m_k\rangle$ when both are expressed in the standard basis. An inspection of Table 1 shows that $|\bar{m}_j\rangle = |m_j\rangle$ for $m = 0,1$ but that $|\bar{m}_j\rangle = |m_{5-j}\rangle$ for $m = 2,3,4$.

The state the physicist must prepare can now be specified. It is

$$|\Psi_0\rangle = \frac{1}{2}\sum_{k=1}^{4}|m_k\rangle|\bar{m}_k\rangle, \quad m = 0,1,2,3 \text{ or } 4$$

$$= \frac{1}{2}(|0000\rangle + |0101\rangle + |1010\rangle + |1111\rangle) \quad . \quad (1)$$

$$= \frac{1}{\sqrt{2}}(|00\rangle + |11\rangle)_{13} \otimes \frac{1}{\sqrt{2}}(|00\rangle + |11\rangle)_{24}$$

The first line shows that this state can be expressed in five alternative ways in terms of the different sets of unbiased bases. The second line shows this state in terms of the standard basis states of the individual qubits, with the first two positions in each ket being for the object qubits and the last two for the ancilla qubits. The last line shows that this state is the tensor product of two Bell states, with one member of each Bell pair (namely, qubits 1 and 2) going to the king and the others (namely, qubits 3 and 4) being retained by the physicist.

The first line of (1) shows that when the king carries out the measurements corresponding to basis $m$ on his qubits and obtains the result $|m_k\rangle$, the state of the entire system collapses into $|m_k\rangle|\bar{m}_k\rangle \equiv |m_k\bar{m}_k\rangle$. The physicist can therefore predict the king's result if she succeeds in finding an orthonormal and complete set of states in the object-ancilla space with the property that each state has a nonvanishing overlap with only a single "doubled" state $|m_k\bar{m}_k\rangle$ from each basis $m$. For she can then measure an observable having the said states as non-degenerate eigenstates and tell, from the result she gets, what eigenvalues the king must have found for the observables he did actually measure.

The construction of the needed object-ancilla states can be accomplished in three steps. The first step consists of introducing an orthonormal basis in the object-ancilla space consisting of $|\Psi_0\rangle$ and the 15 other states

$$\left|\Psi_{(d-1)m+j}\right\rangle = d^{-1/2}\sum_{k=1}^{d}|m_k\bar{m}_k\rangle q^{-jk}, \qquad (2)$$

where $m = 0, 1, 2, ..., d$, $j = 1, 2, ..., d-1$, $q = e^{(2\pi i/d)}$ and $d = 4$ (and the result has been quoted in terms of arbitrary $d$ for later convenience). The orthogonality of these states can be demonstrated by a direct calculation or, more elegantly, by the method in Ref.[1]. The second step consists of introducing the states $|[k_0 k_1 ... k_d]\rangle$ having the property of being orthogonal to all but the states $|0_{k_0}\bar{0}_{k_0}\rangle, |1_{k_1}\bar{1}_{k_1}\rangle, ..., |d_{k_d}\bar{d}_{k_d}\rangle$ of the different bases. These states can be constructed as

$$\left|[k_0 k_1 ... k_d]\right\rangle = \frac{1}{d}\left[|\Psi_0\rangle + \sum_{m=0}^{d}\sum_{j=1}^{d-1}q^{jk_m}\left|\Psi_{(d-1)m+j}\right\rangle\right], \qquad (3)$$

where the result has again been quoted for arbitrary $d$ and can be specialized to the present case by taking $d = 4$.

The third and final step consists of identifying $d^2$ orthonormal states of the form $|[k_0 k_1 ... k_d]\rangle$ that can serve as the non-degenerate eigenstates of the observable the physicist must measure in order to learn the king's results. By measuring this observable and noting the eigenstate she gets, the physicist can tell what eigenvalues must have been found if each of the observable sets had been measured. Of course, the physicist's predictions for the sets that were not measured are completely vacuuous, but her prediction for the one set that was measured turns out to be perfectly on the mark. Now, it can be shown[1] that the condition for two states of the form $|[k_0 k_1 ... k_d]\rangle$ and $|[k_0' k_1' ... k_d']\rangle$ to be orthogonal is that $k_m = k_m'$ for exactly one value of $m$. Thus the physicist is faced with the following arithmetical task in clinching the solution to the king's problem: she must use the numbers from 1 to $d$ to form $d^2$ ordered sets of $(d+1)$ numbers each, with repetitions of numbers within a set allowed, in such a way that any two sets have exactly one identical number in the same place in both. The solution to this task for prime $d$ was given by Englert and Aharonov.[1] We show how a slight modification of their procedure leads to a solution for prime power $d$ as well.

Let $d = p^n$, where $p$ is a prime and $n$ an integer. Our strategy for constructing the state labels in $|k_0 k_1 ... k_d\rangle$ is to allow the "seed" labels $k_0$ and $k_1$ to range independently over the integers $0, 1, ..., d-1$ and fix the remaining labels $k_m$ (for $2 \leq m \leq d$) from the relation

$$k_m = (m-1)k_0 + k_1 \qquad (4)$$

where, however, all numbers on the right are to be reinterpreted as elements of the Galois field $GF(p^n)$ and combined in the manner appropriate to such elements before the result is translated back into an integer in the range $0$ to $d-1$ (note: the number $0$ occuring in the generated state labels should finally be replaced by $d$, in order to conform to the notation for state labels used earlier). This procedure differs from that of Englert and Aharonov[1] only in using $GF(p^n)$, rather than $GF(p)$, to do the arithmetic on the right side of (4). The following paragraph spells out the above recipe in greater detail by explaining what the Galois field $GF(p^n)$ is and how one works with it. The reason this recipe works is to be found in the fact that the elements of $GF(p^n)$ constitute a field.[15]

We give only the briefest account of Galois field theory, stressing just the few ideas needed to perform the calculations in this paper. Highly compressed but useful accounts of Galois fields can be found in Refs.[7] and [16], whereas a truly encyclopedic treatment can be found in Ref.[17]. The Galois field $GF(p)$, with $p$ a prime, is the finite field formed by the numbers $0, 1, 2, ..., p-1$ under addition and multiplication modulo $p$. The "extension" field $GF(p^n)$ is obtained from $GF(p)$ by adjoining to it the root of an irreducible $n$-th degree polynomial with coefficients in $GF(p)$. The adjective "irreducible" refers to the fact that this polynomial cannot be factored in $GF(p)$, i.e that it does not vanish when the unknown in it is set equal to any of the elements of $GF(p)$. Although various choices of the irreducible polynomial are possible, they all give rise to essentially the same field $GF(p^n)$. Let us denote by $\alpha$ the element adjoined to $GF(p)$ to obtain $GF(p^n)$. Then it turns out that the quantities $1, \alpha, \alpha^2, ..., \alpha^{n-1}$ serve as a basis for a vector field over $GF(p)$ whose $p^n$ elements are just the elements of the field $GF(p^n)$. The last remark shows how to set up a correspondence between the numbers $0, 1, 2, ..., d-1$ and the elements of $GF(p^n \equiv d)$: one simply expresses the numbers to base $p$, in the form $x = \sum_{k=0}^{n-1} c_k p^k$, and then replaces the powers $p^k$ by $\alpha^k$.

Arithmetic in the field $GF(p^n)$ is greatly facilitated by the fact that every non-zero element can be expressed as a power of the adjoined element $\alpha$, which is therefore often referred to as a primitive element. Multiplication of elements then reduces to the addition of exponents. Linear combinations of elements can also be replaced by a single element by using the fact that the particular linear combination occuring in the irreducible polynomial vanishes. One can use both of these devices to carry out the arithmetic on the right side of (4) and express the result as a single element of $GF(p^n)$, which can then be converted back into an integer in the range 0 to $d-1$.

We illustrate the above remarks by showing how to solve the king's problem in dimension $d = 2^2$. A suitable second-degree irreducible polynomial over $GF(2)$ that can be used to generate $GF(2^2)$ is $\alpha^2 + \alpha + 1$ (note that neither 0 nor 1 is a root of this polynomial). The numbers 0,1,2 and 3 correspond to the $GF(2^2)$ elements $0 \cdot \alpha + 0 = 0, 0 \cdot \alpha + 1 = 1, 1 \cdot \alpha + 0 = \alpha$ and $1 \cdot \alpha + 1 = \alpha + 1$. From the fact that $\alpha^2 + \alpha + 1 = 0$ and that all polynomial coefficients are defined modulo 2, one can reexpress the numbers 1,2 and 3 as the powers $\alpha^3, \alpha$ and $\alpha^2$ of the primitive element $\alpha$. Using these facts, one can readily work out the addition and multiplication tables for the elements of $GF(2^2)$ and obtain the results shown in Table 2(a). With the aid of these results, one can do the arithmetic on the right side of (4) and obtain the 16 states $|[k_0 k_1 ... k_4]\rangle$, shown in Table 2(b), that solve the king's problem in dimension $2^2$. This solution is essentially unique, upto a permutation of the state labels or a rearrangement in their relative positions.

It is worth adding a few words about some tricks one might like to perform but cannot. Consider the triads of commuting observables $\{1Z, Z1, ZZ\}, \{X1, 1X, XX\}$ and $\{ZX, XZ, YY\}$ that define a set of three mutually unbiased bases. Suppose the king decides to measure one of these triads and challenges the physicist to predict his results. Can she do it? It turns out that she cannot, at least with the method described above, because these three unbiased bases are not part of a maximal set of five. Another challenge the king might pose to the physicist is that he measure any one of the 15 triads of commuting observables that exist and that she predict his results. This overly ambitious trick was shown to be impossible by Mermin.[5] The present trick can be regarded as a generalization of Mermin's successful trick[5] in which the king measures any one of the 15 nontrivial two-qubit observables and challenges the physicist to predict the eigenvalue he finds; our generalization consists of allows the king to measure an entire triad of commuting observables (rather than just one), but limits him to just five specific triads instead of allowing him to range freely over all fifteen.

## 3. Solution in dimension d = $2^3$ = 8.

The king is now given a system of three qubits on which to make his measurements. The sets of commuting observables he is allowed to measure are the ones shown to the left of each row of Table 3.[18] The eigenstates of each set of observables are shown to its right, and the nine orthonormal sets of eigenstates so obtained constitute a set of nine mutually unbiased bases. The notation and conventions for these eigenstates are obvious extensions of those used in Table 1.

In addition to the three "object" qubits given to the king, the physicist retains three "ancilla" qubits in her possession. As before, we denote by $|m_k\rangle$ ($m = 0,1,...,8$ and $k = 1,2,...,8$) the $k$-th state of the $m$-th unbiased basis in the object space and by $|\bar{m}_k\rangle$ the corresponding basis states in the ancilla space. The object states $|m_k\rangle$ are just the ones shown in Table 3, whereas the ancilla states $|\bar{m}_k\rangle$ are chosen to be related to the object states by the condition $\langle 0_j | m_k \rangle = \langle \bar{m}_k | \bar{0}_j \rangle$, which is satisfied by taking $|\bar{0}_j\rangle = |0_j\rangle$ and choosing $|\bar{m}_k\rangle$ to be the complex conjugate of $|m_k\rangle$ in the standard basis. Inspection of Table 3 shows that the $m$-th ancilla basis can always be obtained by suitably permuting states within the $m$-th object basis.

The state of the object and ancilla qubits that the physicist needs to prepare at the start of this trick is

$$|\Psi_0\rangle = \frac{1}{\sqrt{8}} \sum_{k=1}^{8} |m_k\rangle |\bar{m}_k\rangle, \quad m = 0,1,2,3,4,5,6,7 \text{ or } 8, \quad (5)$$

$$= \frac{1}{\sqrt{2}}(|00\rangle + |11\rangle)_{14} \otimes \frac{1}{\sqrt{2}}(|00\rangle + |11\rangle)_{25} \otimes \frac{1}{\sqrt{2}}(|00\rangle + |11\rangle)_{36}, \quad (6)$$

where the first line shows that this state can be expressed in nine alternative ways in terms of the different bases, while the second shows that it is just the tensor product of three Bell states, with one member of each Bell pair going to the king and the other being retained by the physicist (the subscripts 1,..,6 in the second line indicate the relative positions of these qubits in the kets of the first line, when the latter are expanded out in terms of the standard bases of the individual qubits).

The subsequent construction parallels that in Sec.2 and culminates in the task of identifying the 64 orthonormal states $|k_0 k_1 ... k_8\rangle$ that solve the king's problem. This task can be solved using (4) in conjunction with the elements of the Galois field $GF(2^3)$ obtained, for example, by adjoining the root of the irreducible polynomial $\alpha^3 + \alpha + 1$ to $GF(2)$. The solutions for the state labels obtained in this way are shown in Table 4. The remarks made earlier in connection with the $d=4$ solution also apply here: the solution is unique upto a permutation and rearrangement of the state labels, and the king must be restricted to a set of mutually unbiased observables that form part or all of a maximal set if the trick is to be carried out successfully.

## 4. Solution in dimension d = $3^2$ = 9.

This case displays some interesting differences from the earlier two cases because it involves qutrits (i.e. three-state systems) rather than qubits. We begin by introducing the observables for a qutrit[1,10] that play the role of the Pauli operators for a qubit. Denote the standard basis states of a qutrit by $|0\rangle, |1\rangle$ and $|2\rangle$. Then the operators $Z$ and $X$ that generalize the phase- and bit-flip operations on a qubit are defined via their action on the standard basis states as

$$\begin{aligned} Z|0\rangle = |0\rangle, Z|1\rangle = \omega|1\rangle, Z|2\rangle = \omega^2|2\rangle \\ X|0\rangle = |1\rangle, X|1\rangle = |2\rangle, X|2\rangle = |0\rangle \end{aligned} \quad , \quad (7)$$

where $\omega = \exp(2\pi i/3)$ is a cube root of unity. In other words, $Z$ has the standard basis states for its eigenstates while $X$ cyclically permutes these states among themselves. The observables $Z$ and $X$ satisfy the Weyl commutation rule $ZX = \omega XZ$, which is easily seen to be a consequence of (7). In addition to $Z$ and $X$, we will make use of the two further observables $Y \equiv XZ$ and $W \equiv XZ^2$ in the treatment below.[19]

In terms of the above observables, one can construct, using the method of Ref.[10], the ten sets of mutually unbiased observables for a pair of qutrits shown in the first column of Table 5 (each set actually consists of eight commuting observables, but we have just shown a generating pair from which all the others can be constructed as products of powers). The nine simultaneous eigenstates of each set of commuting observables, shown in the second column of Table 5, yield ten mutually unbiased bases for a system of two qutrits. To perform the present trick, the physicist needs to give two "object" qutrits to the king and keep two "ancilla" qutrits in her possession. The observables the king is allowed to measure on his qutrits are the ones shown in the first column of Table 5. Again we introduce mutually unbiased bases $|m_k\rangle$ and $|\bar{m}_k\rangle$ in the object and ancilla spaces, choosing the object bases as in Table 5 and taking each ancilla basis state $|\bar{m}_k\rangle$ to be the complex conjugate of $|m_k\rangle$ in the standard basis. Inspection of Table 5 shows that the ancilla bases are simply shuffled and renamed object bases.

The state prepared by the physicist at the start of this trick is similar to that in the first line of (1) or (6), except for the new bases involved. This state can also be described as the tensor product of two two-qutrit Bell states, with one qutrit of each pair going to the king and the other being retained by the physicist. The rest of the trick proceeds as before and culminates with the physicist having to construct 81 orthonormal object-ancilla states of the form $|[k_0 k_1 ... k_9]\rangle$. This problem can be solved by using (4) in conjunction with the Galois field $GF(3^2)$ obtained by extending $GF(3)$ with the aid of the irreducible polynomial $\alpha^2 + \alpha + 2$. The results for the desired state labels are shown in Table 6.

## 5. Concluding remarks

We have shown how to extend the solution to the king's problem in prime dimensions, given in Ref.[1], to prime power dimensions. Our extension makes crucial use of the construction of mutually unbiased bases given in Ref.[10], but otherwise modifies the solution in Ref.[1] only minimally to accommodate this new case. The construction in Ref.[10] gives a simple answer to one of the questions faced in this work, namely, "What observables should the king be allowed to measure?" However the answer to the question "What observable should the physicist measure in order to learn the king's results?" is not quite as neat. The Englert-Aharonov solution yields the unitary transformation connecting the standard basis to the eigenstates the physicist has to measure, but the task of realizing this transformation through an efficient sequence of one- and two-qubit (or qudit) gates must still be faced before this puzzle can reach the state of experimental realization. However it is encouraging to note that the simplest version of the King's Problem, involving retrodiction of the state of a spin-1/2 particle[2], has recently been realized in a quantum-optical experiment.[20]

The problem of constructing a maximal set of mutually unbiased bases in composite dimensions that are not prime powers, and the solution to the king's problem in these same dimensions, still remain open.

**Acknowledgement.** I am grateful to Berthold-Georg Englert for drawing my attention to an erroneous claim in an earlier version of the paper, and for making several valuable suggestions for improvements in the presentation.

| | | | | |
|---|---|---|---|---|
| Z1,1Z,ZZ | $\|0_1\rangle = 1000$ | $\|0_2\rangle = 0100$ | $\|0_3\rangle = 0010$ | $\|0_4\rangle = 0001$ |
| X1,1X,XX | $\|1_1\rangle = 1111$ | $\|1_2\rangle = 1\bar{1}1\bar{1}$ | $\|1_3\rangle = 11\bar{1}\bar{1}$ | $\|1_4\rangle = 1\bar{1}\bar{1}1$ |
| Y1,1Y,YY | $\|2_1\rangle = 1ii\bar{1}$ | $\|2_2\rangle = 1\bar{i}i1$ | $\|2_3\rangle = 1i\bar{i}1$ | $\|2_4\rangle = 1\bar{i}\,\bar{i}\,\bar{1}$ |
| XY,YZ,ZX | $\|3_1\rangle = 1\bar{1}ii$ | $\|3_2\rangle = 11\bar{i}i$ | $\|3_3\rangle = 11i\bar{i}$ | $\|3_4\rangle = 1\bar{1}\bar{i}\,\bar{i}$ |
| YX,ZY,XZ | $\|4_1\rangle = 1i\bar{1}i$ | $\|4_2\rangle = 1\bar{i}1i$ | $\|4_3\rangle = 1i1\bar{i}$ | $\|4_4\rangle = 1\bar{i}\,\bar{1}\,\bar{i}$ |

TABLE 1. Mutually unbiased bases for a system of two qubits. Each row shows the four simultaneous eigenstates of the three commuting observables to the left, arranged according to the eigenvalue signatures $++, +-, -+$ and $--$ with respect to the first two of these observables. The eigenstates in any two rows are mutually unbiased in the sense that the squared modulus of the inner product between any two of them (one from each row) is the same and equal to ¼. The numbers *abcd* after each ket are used as a shorthand for the (unnormalized) state $a|00\rangle + b|01\rangle + c|10\rangle + d|11\rangle$. Note: $i = \sqrt{-1}$ and a bar over a number indicates its negative.

| + | 0 | 1 | $\alpha$ | $\alpha+1$ |
|---|---|---|---|---|
| 0 | 0 | 1 | $\alpha$ | $\alpha+1$ |
| 1 | 1 | 0 | $\alpha+1$ | $\alpha$ |
| $\alpha$ | $\alpha$ | $\alpha+1$ | 0 | 1 |
| $\alpha+1$ | $\alpha+1$ | $\alpha$ | 1 | 0 |

| × | 0 | 1 | $\alpha$ | $\alpha+1$ |
|---|---|---|---|---|
| 0 | 0 | 0 | 0 | 0 |
| 1 | 0 | 1 | $\alpha$ | $\alpha+1$ |
| $\alpha$ | 0 | $\alpha$ | $\alpha+1$ | 1 |
| $\alpha+1$ | 0 | $\alpha+1$ | 1 | $\alpha$ |

TABLE 2(a). Addition and multiplication tables for the elements of $GF(2^2)$.

```
11432   12341   13214   14123
21324   22413   23142   24231
31243   32134   33421   34312
41111   42222   43333   44444
```

TABLE 2(b). State labels of the 16 states $|[k_0 k_1 ... k_4]\rangle$ that solve the king's problem in dimension $2^2 = 4$.

| | | | | |
|---|---|---|---|---|
| Z11,1Z1,11Z | $\lvert 0_1\rangle = 10000000$ | $\lvert 0_2\rangle = 01000000$ | $\lvert 0_3\rangle = 00100000$ | $\lvert 0_4\rangle = 00010000$ |
| | $\lvert 0_5\rangle = 00001000$ | $\lvert 0_6\rangle = 00000100$ | $\lvert 0_7\rangle = 00000010$ | $\lvert 0_8\rangle = 00000001$ |
| X11,1X1,11X | $\lvert 1_1\rangle = 11111111$ | $\lvert 1_2\rangle = 1\bar{1}1\bar{1}1\bar{1}1\bar{1}$ | $\lvert 1_3\rangle = 11\bar{1}\bar{1}11\bar{1}\bar{1}$ | $\lvert 1_4\rangle = 1\bar{1}\bar{1}11\bar{1}\bar{1}1$ |
| | $\lvert 1_5\rangle = 1111\bar{1}\bar{1}\bar{1}\bar{1}$ | $\lvert 1_6\rangle = 1\bar{1}1\bar{1}\bar{1}1\bar{1}1$ | $\lvert 1_7\rangle = 11\bar{1}\bar{1}\bar{1}\bar{1}11$ | $\lvert 1_8\rangle = 1\bar{1}\bar{1}1\bar{1}11\bar{1}$ |
| Y11,1Y1,11Y | $\lvert 2_1\rangle = 1ii\bar{1}i\bar{1}\bar{1}\bar{i}$ | $\lvert 2_2\rangle = 1\bar{i}i1i1\bar{1}i$ | $\lvert 2_3\rangle = 1i\bar{i}1i\bar{1}1i$ | $\lvert 2_4\rangle = 1\bar{i}\,\bar{i}\,\bar{1}i11\bar{i}$ |
| | $\lvert 2_1\rangle = 1ii\bar{1}i\bar{1}\bar{1}\bar{i}$ | $\lvert 2_2\rangle = 1\bar{i}i1i1\bar{1}i$ | $\lvert 2_3\rangle = 1\,i\bar{i}\,1i\bar{1}1i$ | $\lvert 2_4\rangle = 1\bar{i}\,\bar{i}\,\bar{1}i11\bar{i}$ |
| XYX,XZZ,YYZ | $\lvert 3_1\rangle = 1\,i1\,i1\bar{i}\,\bar{1}\,i$ | $\lvert 3_2\rangle = 1\bar{i}\,\bar{1}i1i1i$ | $\lvert 3_3\rangle = 1\,i\bar{1}\,\bar{i}\,\bar{1}i\bar{1}i$ | $\lvert 3_4\rangle = 1\bar{i}1\bar{i}\,\bar{1}\bar{i}1i$ |
| | $\lvert 3_5\rangle = 1\bar{i}1\bar{i}1i\bar{1}\bar{i}$ | $\lvert 3_6\rangle = 1i\bar{1}\bar{i}\,1\bar{i}\,1\bar{i}$ | $\lvert 3_7\rangle = 1\bar{i}\,\bar{1}\,i\bar{1}\,\bar{i}\,\bar{1}\,\bar{i}$ | $\lvert 3_8\rangle = 1\,i1\,i\bar{1}\,i1\,\bar{i}$ |
| XXZ,YXY,YZZ | $\lvert 4_1\rangle = 11\,iii\bar{i}\,1\bar{1}$ | $\lvert 4_2\rangle = 11\bar{i}\,\bar{i}\,\bar{i}i1\bar{1}$ | $\lvert 4_3\rangle = 1\bar{1}\,i\bar{i}ii11$ | $\lvert 4_4\rangle = 1\bar{1}\,\bar{i}i\bar{i}\,\bar{i}11$ |
| | $\lvert 4_5\rangle = 1\,\bar{1}\,\bar{i}\,ii\,i\bar{1}\,\bar{1}$ | $\lvert 4_6\rangle = 1\,\bar{1}\,\bar{i}\,ii\,i\bar{1}\,\bar{1}$ | $\lvert 4_7\rangle = 11\bar{i}\,\bar{i}i\,\bar{i}\,\bar{1}1$ | $\lvert 4_8\rangle = 11\,i\,i\,\bar{i}\,i\bar{1}1$ |
| YXX,YZY,ZZX | $\lvert 5_1\rangle = 11\,i\bar{i}1\bar{1}ii$ | $\lvert 5_2\rangle = 1\bar{1}\,i\,i\bar{1}\,\bar{1}\,\bar{i}\,i$ | $\lvert 5_3\rangle = 11\bar{i}\,i\bar{1}1i\,i$ | $\lvert 5_4\rangle = 1\bar{1}\,\bar{i}\,\bar{i}11\bar{i}\,i$ |
| | $\lvert 5_5\rangle = 11\bar{i}\,i1\bar{1}\,\bar{i}\,\bar{i}$ | $\lvert 5_6\rangle = 1\bar{1}\,\bar{i}\,\bar{i}\,\bar{1}\bar{1}i\bar{i}$ | $\lvert 5_7\rangle = 11i\bar{i}\,\bar{1}1\bar{i}\,\bar{i}$ | $\lvert 5_8\rangle = 1\bar{1}\,i\,i11i\,\bar{i}$ |
| YYX,ZXX,ZYZ | $\lvert 6_1\rangle = 1ii11i\bar{i}\,\bar{1}$ | $\lvert 6_2\rangle = 1\bar{i}\,\bar{i}11\bar{i}\,i\bar{1}$ | $\lvert 6_3\rangle = 1\bar{i}\,i\bar{1}\bar{1}ii\bar{1}$ | $\lvert 6_4\rangle = 1i\bar{i}\,\bar{1}\bar{1}\bar{i}\,\bar{i}\,\bar{1}$ |
| | $\lvert 6_5\rangle = 1ii1\bar{1}\,\bar{i}\,i1$ | $\lvert 6_6\rangle = 1\bar{i}\,\bar{i}1\bar{1}i\bar{i}1$ | $\lvert 6_7\rangle = 1\bar{i}\,i\bar{1}1\bar{i}\,\bar{i}1$ | $\lvert 6_8\rangle = 1\,i\,\bar{i}\,\bar{1}1\,i\,i1$ |
| XYY,XZX,ZZY | $\lvert 7_1\rangle = 1i1\bar{i}\,i1i\bar{1}$ | $\lvert 7_2\rangle = 1\bar{i}1i\bar{i}1\bar{i}\,\bar{1}$ | $\lvert 7_3\rangle = 1i\bar{1}i\bar{i}\,\bar{1}i\bar{1}$ | $\lvert 7_4\rangle = 1\bar{i}\,\bar{1}\,\bar{i}\,i\bar{1}\,\bar{i}\,\bar{1}$ |
| | $\lvert 7_5\rangle = 1i\bar{1}\,ii1\bar{i}1$ | $\lvert 7_6\rangle = 1\bar{i}\,\bar{1}\,\bar{i}\,\bar{i}1i1$ | $\lvert 7_7\rangle = 1i1\bar{i}\,\bar{i}\,\bar{1}\bar{i}1$ | $\lvert 7_8\rangle = 1\bar{i}1i\,i\bar{1}i1$ |
| XXY,ZXZ,ZYY | $\lvert 8_1\rangle = 111\bar{1}ii\,\bar{i}\,i$ | $\lvert 8_2\rangle = 1\bar{1}11\bar{i}\,i\,i\,i$ | $\lvert 8_3\rangle = 1\bar{1}11i\bar{i}ii$ | $\lvert 8_4\rangle = 11\bar{1}\bar{1}\bar{i}\,\bar{i}\,\bar{i}\,i$ |
| | $\lvert 8_5\rangle = 111\bar{1}\bar{i}\,\bar{i}\,i\bar{i}$ | $\lvert 8_6\rangle = 1\bar{1}11\bar{i}\,\bar{i}\,\bar{i}\,i$ | $\lvert 8_7\rangle = 1\bar{1}\bar{1}\bar{1}\bar{i}i\bar{i}\,\bar{i}$ | $\lvert 8_8\rangle = 11\bar{1}\bar{1}i\,i\,i\bar{i}$ |

TABLE 3. Mutually unbiased bases for a system of three qubits. Each row shows the eight simultaneous eigenstates of the three commuting observables to the left of that row, with the eigenstates ordered according to the eigenvalue signatures $+++,++-,...,---$. The eigenstates in any two rows are mutually unbiased in the sense that the squared modulus of the inner product between any two of them (one from each row) is the same and equal to 1/8. The numbers *abcdefgh* following any ket are used as a shorthand for the (unnormalized) state $a\lvert 000\rangle + b\lvert 001\rangle + ... + h\lvert 111\rangle$. Note: $i = \sqrt{-1}$ and a bar over a number indicates its negative.

| | | | | | | | |
|---|---|---|---|---|---|---|---|
| 118325476 | 123816745 | 132187654 | 145678123 | 154761832 | 167452381 | 176543218 | 181234567 |
| 213572064 | 228641357 | 231758246 | 246827531 | 257136428 | 264285713 | 275314682 | 282463175 |
| 312746583 | 321475638 | 338564721 | 347213856 | 356382147 | 365831274 | 374128365 | 383657412 |
| 415267348 | 426154873 | 437845162 | 448732615 | 451623784 | 462518437 | 473481526 | 484376251 |
| 514853627 | 527368514 | 536271485 | 541586372 | 558417263 | 563724158 | 572635841 | 585142736 |
| 617684235 | 624537186 | 635426817 | 642351768 | 653248671 | 668173542 | 671862453 | 686715324 |
| 716438752 | 725783461 | 734612578 | 743165287 | 752874316 | 761347825 | 778256134 | 787521643 |
| 81111111 | 82222222 | 83333333 | 84444444 | 85555555 | 86666666 | 87777777 | 88888888 |

TABLE 4. State labels of the 64 states $|[k_0 k_1 ... k_8]\rangle$ that solve the king's problem in dimension $2^3 = 8$.

| | | | |
|---|---|---|---|
| Z1,1Z | $\|0_1\rangle = 100000000$ | $\|0_2\rangle = 010000000$ | $\|0_3\rangle = 001000000$ |
| | $\|0_4\rangle = 000100000$ | $\|0_5\rangle = 000010000$ | $\|0_6\rangle = 000001000$ |
| | $\|0_7\rangle = 000000100$ | $\|0_8\rangle = 000000010$ | $\|0_9\rangle = 000000001$ |
| X1,1X | $\|1_1\rangle = 1\bar{\omega}\omega\bar{\omega}\omega 1\omega 1\bar{\omega}$ | $\|1_2\rangle = 1\omega\bar{\omega}\bar{\omega}1\omega\omega\bar{\omega}1$ | $\|1_3\rangle = 111\bar{\omega}\bar{\omega}\bar{\omega}\omega\omega\omega$ |
| | $\|1_4\rangle = 1\bar{\omega}\omega\omega 1\bar{\omega}\bar{\omega}\omega 1$ | $\|1_5\rangle = 1\omega\bar{\omega}\bar{\omega}1\bar{\omega}1\omega$ | $\|1_6\rangle = 111\omega\omega\omega\bar{\omega}\bar{\omega}\bar{\omega}$ |
| | $\|1_7\rangle = 1\bar{\omega}\omega 1\bar{\omega}\omega 1\bar{\omega}\omega$ | $\|1_8\rangle = 1\omega\bar{\omega}1\omega\bar{\omega}1\omega\bar{\omega}$ | $\|1_9\rangle = 111111111$ |
| Y1,1Y | $\|2_1\rangle = 1\bar{\omega}\bar{\omega}\bar{\omega}\omega\omega\bar{\omega}\omega\omega$ | $\|2_2\rangle = 1\omega 1\bar{\omega}1\bar{\omega}\bar{\omega}1\bar{\omega}$ | $\|2_3\rangle = 11\omega\bar{\omega}\bar{\omega}1\bar{\omega}\bar{\omega}1$ |
| | $\|2_4\rangle = 1\bar{\omega}\bar{\omega}\omega 11\bar{\omega}\bar{\omega}$ | $\|2_5\rangle = 1\omega 1\omega\bar{\omega}\bar{\omega}1\omega 1$ | $\|2_6\rangle = 11\omega\omega\omega\bar{\omega}11\omega$ |
| | $\|2_7\rangle = 1\bar{\omega}\bar{\omega}1\bar{\omega}\bar{\omega}\omega 11$ | $\|2_8\rangle = 1\omega 11\omega 1\omega\bar{\omega}\omega$ | $\|2_9\rangle = 11\omega 11\omega\omega\omega\bar{\omega}$ |
| W1,1W | $\|3_1\rangle = 1\bar{\omega}1\bar{\omega}\omega\bar{\omega}1\bar{\omega}1$ | $\|3_2\rangle = 1\omega\omega\bar{\omega}111\omega\omega$ | $\|3_3\rangle = 11\bar{\omega}\bar{\omega}\bar{\omega}\omega 11\bar{\omega}$ |
| | $\|3_4\rangle = 11\bar{\omega}\bar{\omega}\bar{\omega}\omega 11\bar{\omega}$ | $\|3_5\rangle = 1\bar{\omega}1\omega 1\omega\omega 1\omega$ | $\|3_6\rangle = 11\bar{\omega}\omega\omega 1\omega\omega 1$ |
| | $\|3_7\rangle = 1\bar{\omega}11\bar{\omega}1\bar{\omega}\omega\bar{\omega}$ | $\|3_8\rangle = 1\omega\omega 1\omega\omega\bar{\omega}11$ | $\|3_9\rangle = 11\bar{\omega}11\bar{\omega}\bar{\omega}\bar{\omega}\omega$ |
| XZ,ZW | $\|4_1\rangle = 1\bar{\omega}1\bar{\omega}\bar{\omega}\omega\omega\bar{\omega}\bar{\omega}$ | $\|4_2\rangle = 1\omega\omega\bar{\omega}\omega\bar{\omega}\omega\omega 1$ | $\|4_3\rangle = 11\bar{\omega}\bar{\omega}11\omega 1\omega$ |
| | $\|4_4\rangle = 1\bar{\omega}1\omega\omega 1\bar{\omega}11$ | $\|4_5\rangle = 1\omega\omega\omega 1\omega\bar{\omega}\bar{\omega}\omega$ | $\|4_6\rangle = 11\bar{\omega}\omega\bar{\omega}\bar{\omega}\bar{\omega}\omega\bar{\omega}$ |
| | $\|4_7\rangle = 1\bar{\omega}111\bar{\omega}1\omega\omega$ | $\|4_8\rangle = 1\omega\omega 1\bar{\omega}11 1\bar{\omega}$ | $\|4_9\rangle = 11\bar{\omega}1\omega\omega 1\bar{\omega}1$ |
| YZ,ZX | $\|5_1\rangle = 1\bar{\omega}\omega\bar{\omega}\bar{\omega}\bar{\omega}\bar{\omega}1\omega$ | $\|5_2\rangle = 1\omega\bar{\omega}\bar{\omega}\omega 1\bar{\omega}\bar{\omega}\bar{\omega}$ | $\|5_3\rangle = 111\bar{\omega}1\omega\bar{\omega}\omega 1$ |
| | $\|5_4\rangle = 1\bar{\omega}\omega\omega\omega\omega 1\omega\bar{\omega}$ | $\|5_5\rangle = 1\omega\bar{\omega}\omega 1\bar{\omega}111$ | $\|5_6\rangle = 111\omega\omega\bar{\omega}11\bar{\omega}\omega$ |
| | $\|5_7\rangle = 1\bar{\omega}\omega 111\omega\bar{\omega}1$ | $\|5_8\rangle = 1\omega\bar{\omega}1\bar{\omega}\omega\omega\omega\omega\omega$ | $\|5_9\rangle = 1111\omega\bar{\omega}\bar{\omega}\omega 1\bar{\omega}$ |
| $XZ^2,Z^2Y$ | $\|6_1\rangle = 1\bar{\omega}\bar{\omega}\bar{\omega}1\bar{\omega}\omega\omega\bar{\omega}$ | $\|6_2\rangle = 1\omega 1\bar{\omega}\bar{\omega}1\omega 11$ | $\|6_3\rangle = 11\omega\bar{\omega}\bar{\omega}\omega\omega\omega\bar{\omega}\omega$ |
| | $\|6_4\rangle = 1\bar{\omega}\bar{\omega}\omega\bar{\omega}\omega\bar{\omega}\bar{\omega}1$ | $\|6_5\rangle = 1\omega 1\omega\omega\bar{\omega}\bar{\omega}\omega\omega$ | $\|6_6\rangle = 11\omega\omega 11\bar{\omega}1\bar{\omega}$ |
| | $\|6_7\rangle = 1\bar{\omega}\bar{\omega}1\omega 111\omega$ | $\|6_8\rangle = 1\omega 111\omega 1\bar{\omega}\bar{\omega}$ | $\|6_9\rangle = 11\omega 1\bar{\omega}\bar{\omega}1\omega 1$ |
| $YZ^2,Z^2W$ | $\|7_1\rangle = 1\bar{\omega}1\bar{\omega}11\bar{\omega}\bar{\omega}\omega$ | $\|7_2\rangle = 1\omega\omega\bar{\omega}\bar{\omega}\omega\omega\bar{\omega}\omega\bar{\omega}$ | $\|7_3\rangle = 11\bar{\omega}\bar{\omega}\omega\omega\bar{\omega}\bar{\omega}11$ |
| | $\|7_4\rangle = 1\bar{\omega}1\omega\bar{\omega}\bar{\omega}11\bar{\omega}$ | $\|7_5\rangle = 1\omega\omega\omega\omega 11\bar{\omega}1$ | $\|7_6\rangle = 11\bar{\omega}\omega 1\omega 1\omega\omega$ |
| | $\|7_7\rangle = 1\bar{\omega}11\omega\omega\omega\omega 1$ | $\|7_8\rangle = 1\omega\omega 11\bar{\omega}\omega 1\omega$ | $\|7_9\rangle = 11\bar{\omega}1\bar{\omega}1\omega\bar{\omega}\bar{\omega}$ |
| WZ,ZY | $\|8_1\rangle = 1\bar{\omega}\bar{\omega}\bar{\omega}\bar{\omega}11\omega 1$ | $\|8_2\rangle = 1\omega 1\bar{\omega}\omega\omega 11\omega$ | $\|8_3\rangle = 11\omega\omega 1\bar{\omega}1\bar{\omega}\omega$ |
| | $\|8_4\rangle = 1\bar{\omega}\bar{\omega}\omega\omega\bar{\omega}\omega\bar{\omega}\omega$ | $\|8_5\rangle = 1\omega 1\omega 11\omega\omega\bar{\omega}$ | $\|8_6\rangle = 11\omega\omega\bar{\omega}\omega\omega 11$ |
| | $\|8_7\rangle = 1\bar{\omega}\bar{\omega}11\omega\bar{\omega}1\bar{\omega}$ | $\|8_8\rangle = 1\omega 11\bar{\omega}\bar{\omega}\bar{\omega}\bar{\omega}1$ | $\|8_9\rangle = 11\omega 1\omega 1\bar{\omega}\omega\omega$ |
| $WZ^2,Z^2X$ | $\|9_1\rangle = 1\bar{\omega}\omega\bar{\omega}1\omega 111$ | $\|9_2\rangle = 1\omega\bar{\omega}\bar{\omega}\bar{\omega}\bar{\omega}1\bar{\omega}\omega$ | $\|9_3\rangle = 111\bar{\omega}\omega 11\omega\bar{\omega}$ |
| | $\|9_4\rangle = 1\bar{\omega}\omega\omega\bar{\omega}1\omega\omega\omega$ | $\|9_5\rangle = 1\omega\bar{\omega}\omega\omega\omega\omega 1\bar{\omega}$ | $\|9_6\rangle = 111\omega 1\bar{\omega}\omega\bar{\omega}1$ |
| | $\|9_7\rangle = 1\bar{\omega}\omega 1\omega\bar{\omega}\bar{\omega}\bar{\omega}\bar{\omega}$ | $\|9_8\rangle = 1\omega\bar{\omega}111\bar{\omega}\omega 1$ | $\|9_9\rangle = 1111\bar{\omega}\omega\bar{\omega}1\omega$ |

(caption for table on previous page)
TABLE 5. Mutually unbiased bases for a system of two qutrits. Each row shows the nine simultaneous eigenstates of the two commuting observables to the left, with $\omega = \exp(2\pi i/3)$ and $\bar{\omega} = \exp(-2\pi i/3)$ (and not $-\omega$). The numbers *abcdefghi* following any ket are used as a shorthand for the (unnormalized) state $a|00\rangle + b|01\rangle + ... + i|22\rangle$, with $0, 1$ and $2$ referring to the standard basis states of a qutrit. The eigenstates in any row are ordered so that they have the eigenvalue signatures $\omega\omega, \omega\bar{\omega}, \omega 1, ..., 1\bar{\omega}, 11$ with respect to the two observables on the left. The eigenstates in any two rows are mutually unbiased in the sense that the squared modulus of the inner product between any two of them is the same and equal to 1/9.

| | | | | | | |
|---|---|---|---|---|---|---|
| 1129453786 | 1291534867 | 1345678912 | 1453786129 | 1534867291 | 1678912345 | 1786129453 |
| 1867291534 | 1912345678 | 2192768435 | 2219876543 | 2354921687 | 2435192768 | 2543219876 |
| 2687354921 | 2768435192 | 2876543219 | 2921687354 | 3147825369 | 3258693471 | 3369147825 |
| 3471258693 | 3582936714 | 3693471258 | 3714582936 | 3825369147 | 3936714582 | 4156237948 |
| 4237948156 | 4372489561 | 4489561372 | 4561372489 | 4615723894 | 4723894615 | 4894615723 |
| 4948156237 | 5138579624 | 5246381795 | 5381795246 | 5462813957 | 5579624138 | 5624138579 |
| 5795246381 | 5813957462 | 5957462813 | 6174396852 | 6285417639 | 6396852174 | 6417639285 |
| 6528741963 | 6639285417 | 6741963528 | 6852174396 | 6963528741 | 7183642507 | 7264759318 |
| 7318264759 | 7426975831 | 7597183642 | 7642597183 | 7759318264 | 7831426975 | 7975831426 |
| 8165984273 | 8273165984 | 8327516498 | 8498327516 | 8516498327 | 8651849732 | 8732651849 |
| 8849732651 | 8984273165 | 9111111111 | 9222222222 | 9333333333 | 9444444444 | 9555555555 |
| 9666666666 | 9777777777 | 9888888888 | 9999999999 | | | |

TABLE 6. State labels of the 81 states $|[k_0 k_1 ... k_9]\rangle$ that solve the king's problem in dimension $3^2 = 9$.